# GRB duration distribution considering the position of the Fermi

Dorottya Szécsi[1], Zsolt Bagoly[1,2], István Horváth[2], Lajos G. Balázs[3], Péter Veres[1,2,3], Attila Mészáros[4]

[1]*Eötvös University, Budapest*
[2]*Bolyai Military University, Budapest*
[3]*Konkoly Observatory, Budapest*
[4]*Charles University, Prague*

**Abstract**

The Fermi satellite has a particular motion during its flight which enables it to catch the gamma-ray bursts mostly well. The side-effect of this favourable feature is that the lightcurves of the GBM detectors are stressed by rapidly and extremely varying background. Before this data is processed, it needs to be separated from the background. The commonly used methods [3], [7] were useless for most cases of Fermi, so we developed a new technique based on the motion and orientation of the satellite. The background-free lightcurve can be used to perform statistical surveys, hence we showed the efficiency of our background-filtering method presenting a statistical analysis known from the literature.

## 1   Introduction

NASA's Fermi Gamma-ray Space Telescope is designed to measure the position of a burst in seconds and change the detectors' orientation so that the orientation of the effective area of the main detector (LAT) and the celestial position of the burst have the smallest angle [6]. The problem is that the secondary detector (GBM) measures a quickly varying background superposed on the substantial data because of the rapid motion of the satellite. Though, work with the GBM data must be anticipated by removing the background, the widely-used methods, like fitting the background with a polynomial function of time are too simple and not very effective in the case of Fermi. We made a model that involves the real motion and orientation of the satellite during its flight and the position and contribution of the main gamma-ray sources to the background.

## 2   The data and the model

After some basic data transformation, we can plot a GBM lightcurve as shown in Fig. 1, which is typical.

Since the variation of the background is caused by the rapid motion of the satellite, we need a specific model involving the position and orientation of the Fermi in every second. We made such a model considering the detector's orientation and the celestial position of the burst, the Sun and the Earth limb. The effect of every other gamma-source is included in an isotropic constant gamma-ray background in this approximation. If the Earth-limb is in the detector's field of view, it has two effects. First, it shields the cosmic gamma-ray background, which in our case we presume to be homogeneous and isotropic. Second, there are so-called terrestric gamma-ray sources, like lightning in the upper atmosphere. These sources can influence the actual level of the background. [6] In both cases, the detected background depends on how much the Earth-limb shields from the detector's field of view. In order to measure this, we compute the rate of the uncovered sky correlated to the size of the field of view and use this as an underlying variable.

Comparing Fig. 1 to Fig. 2, we can note the connection between them: where the angles and the Earth limb's uncovering vary, the background of the lightcurve also varies. Therefore, we use these 3 variables to fit the background of the burst.

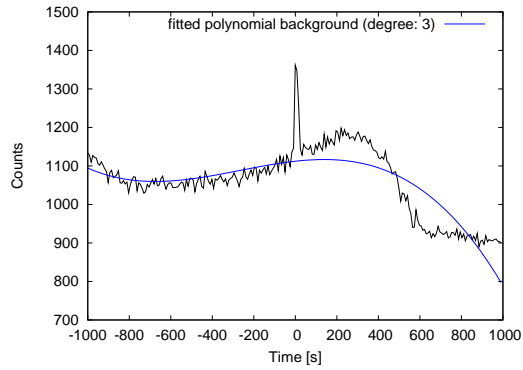

Figure 1: Lightcurve of the Fermi-burst 091030613 measured by the 3rd GBM detector. The burst is at 0 s but the variation of the background is comparable with its height, and cannot be modelled by a simple polynomial function of time of degree 2 or 3.

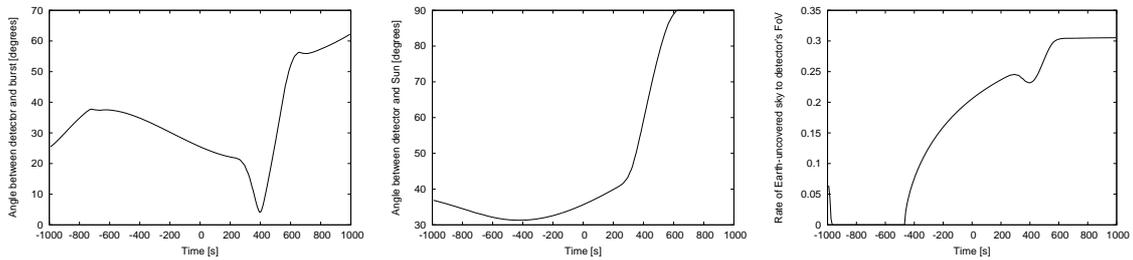

Figure 2: *From left to right:* The celestial distance of the 3rd GBM detector and the position of Fermi-burst 091030613; the same with the Sun; and the rate of the Earth-uncovered sky to the 3rd GBM detector's field of view, both in the function of time. It is worth comparing these figures with Fig. 1.

## 3  Fitting the background

In order to separate the background from the burst's real data, we fitted a 3-dimensional hypersurface of degree 3 to the lightcurve's background. The 3 dimensions were shown in Fig. 2. After subtracting the hypersurface from the lightcurve, we get the background-free lightcurve shown in Fig. 3.

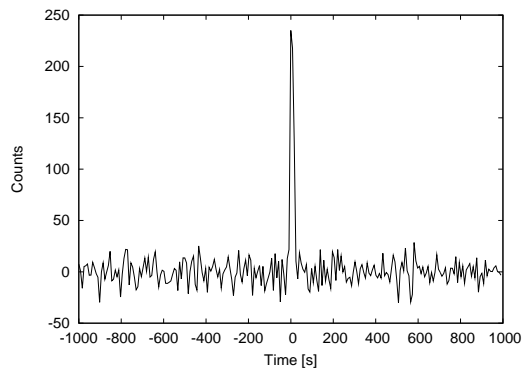

Figure 3: After background fitting: the lightcurve of Fermi-burst 091030613 measured by the 3rd GBM detector. Comparing Fig. 1 to Fig. 3, we can state that our background filtering was successful, since the resulting lightcurve is perfectly background-free.

# 4 Testing the method

In order to verify our method described above, we made a statistical analysis based on our background-filtered data set. Since we were investigating the temporal characteristics of the lightcurve, it was obvious to compute the $T_{90}$ statistical parameter (see [3]). We ran our background filtering method on 332 Fermi-bursts and computed the $T_{90}$ values for them. It is well-known from the literature that the (logarithmic) distribution of the gamma-ray bursts' duration shows 2 or 3 peaks (see for example [1] or [4]). We present the distribution of the $\log T_{90}$ variable computed by us in Fig. 4.

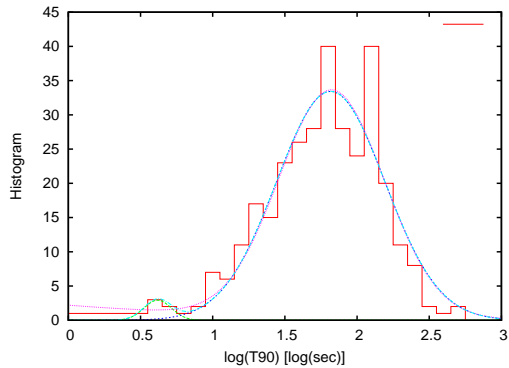

Figure 4: The distribution of the $\log T_{90}$ statistical parameter for 332 bursts according to our method. The coloured curves show the fitted Gaussian functions: green and blue for the short peak and the long peak, light blue and purple for together.

In Fig. 4, the duration distribution of our data follows the general $T_{90}$ distribution shape (e.g. it has two peaks). The greater peak on the right-hand side surely represents the long/soft bursts, while the smaller peak on the left-hand side can represent the short/hard bursts or also the intermediate group (see [2] and [5]). We need further investigations to decide this question, but since the shape of the distribution is analogous to the literature (logarithmic distribution with two peaks), we will develop and use this method for further Fermi GRB analyses.

# 5 Conclusion

The method can be further developed by the effects of other gamma-ray sources, e.g. the Moon, some galactical sources or near supernova remnants. Spectral (energy dependent) analysis will be performed as well, which is one of our goals in the future. However, we can announce that we have succesfully created a method that is able to separate the Fermi data from the motion-based background in such an effective way that statistical analyses can be made using the data.

## Acknowledgment

This work was supported by OTKA grant K077795, by OTKA/NKTH A08-77719 and A08-77815 grants (Z.B.), by GAČR grant No. P209/10/0734 (A.M.), by the Research Program MSM0021620860 of the Ministry of Education of the Czech Republic (A.M.) and by a Bolyai Scholarship (I.H.).